\documentclass[prl,twocolumn]{revtex4}

\usepackage{graphicx}
\usepackage{dcolumn}
\usepackage{bm}

\begin{document}

\preprint{cond-mat/???????}

\title{Imaging density disturbances in water with 41.3 attosecond time resolution}%

\author{P. Abbamonte$^1$}
 \altaffiliation{Current address: Bldg. 725D, Brookhaven National Laboratory, Upton, NY, 11973}
\author{K. D. Finkelstein$^2$}%
\author{M. D. Collins$^1$}%
\author{S. M. Gruner$^{1,2}$}%
\affiliation{$^1$Department of Physics, $^2$Cornell High Energy Synchrotron Source, Cornell University, Ithaca, NY, 14853-2501}%

\date{\today}

\begin{abstract}
We show that the momentum flexibility of inelastic x-ray scattering may be exploited to invert its loss
function, alowing real time imaging of density disturbances in a medium.  
We show the disturbance arising from a point 
source in liquid water, with a resolution of 41.3 attoseconds ($4.13 \times 10^{-17}$ sec) and 1.27 $\AA$ 
($1.27 \times 10^{-8}$ cm).  This result is used to determine the structure of the electron cloud 
around a photoexcited 
chromophore in solution, as well as the wake generated in water by a 9 MeV gold ion.  
We draw an analogy with pump-probe techniques and suggest that energy-loss 
scattering may be applied more generally to the study of attosecond phenomena.
\end{abstract}

\maketitle

Brisk progress has been made recently in the generation and 
detection of ultrashort, attosecond (1 as = 10$^{-18}$ sec) 
laser pulses with high harmonic generation techniques
\cite{drescher1,drescher2,baltiuska,hentschel,keinberger,niikura1,niikura2,paul}.  
This has heralded an age of attophysics in which many-electron dynamics will be probed in real time.  
Much of what is already known about electron dynamics has been 
derived from energy-loss techniques such as inelastic x-ray, electron, or neutron scattering, 
which have been 
fruitfully applied to the study of, for example, plasma oscillations, exciton dynamics, and spin waves.  
Such experiments are normally examined in the frequency / momentum representation, where the spectra may be
easily compared to theoretically calculable response functions for the electron density or local 
magnetic moment.  However if such measurements truly reflect dynamics, a temporal representation
may also be illuminating.

In this letter we demonstrate a method for inverting energy-loss measurements into
time and space, permitting explicit imaging of electron dynamics in a medium.  This allows
the spatial extent of excited states to be determined, i.e. the mean free path or dephasing distance, 
and a parallel with pump-probe techniques to be drawn.
Inverting requires reliable sampling of a loss 
function along both energy and momentum axes with sufficient range to resolve the features of 
interest and sufficient resolution to provide an adequate field of view.  Therefore, 
because of its kinematic flexibility, we consider the  
case of inelastic x-ray scattering (IXS).

In IXS a photon with well-defined initial momentum and energy, 
$({\bf k}_i,\omega_i)$, is impinged on a specimin which scatters it to a final $({\bf k}_f,\omega_f)$.  
The spectral density 
of scattered photons is proportional to the dynamic structure factor of the material, 
$S({\bf k},\omega)$, where 
$\omega = \omega_i - \omega_f$  and ${\bf k} = {\bf k}_i - {\bf k}_f$ are the transferred 
energy and momentum\cite{schuelke}.  $S({\bf k},\omega)$ was posed 
originally by van Hove\cite{vanHove} as a measure of the dynamical properties 
of an interacting electron system, and 
by the fluctuation-dissipation theorem is related to the imaginary part of a response function\cite{sturm}

\begin{equation}
Im \left [\chi ({\bf k},\omega) \right ] = - \pi \left [ S({\bf k},\omega) - S({\bf k},-\omega) \right ]
\end{equation}

\noindent
which can thereby be experimentally determined.  
$\chi({\bf k},\omega)$ is known as the density Green's function, or density propagator, and describes the way disturbances 
in the electron density propagate in the system.  $\chi({\bf k},\omega)$ is the space-time fourier 
transform of 
$\chi({\bf x},t) = - i <0|[\delta \hat{n} ({\bf x},t),\delta \hat{n} (0,0)]|0> \theta (t)$, 
where $\delta \hat{n}$ is the density fluctuation operator, and in real 
space represents the disturbance produced by a delta function source at the origin at $t = 0$.  Energy 
loss scattering, in this sense, is rather like a pump-probe experiment; the system is perturbed at a 
reference point in time, and its subsequent evolution observed.  To make the analogy more concrete we wish
to carry out such an inversion on a real IXS data set.

Unfortunately, inverting requires knowledge of the full $\chi({\bf k},\omega)$, 
but the experiment provides only its 
imaginary part.  In other words in IXS there is a phase problem, like that in x-ray 
crystallography, that must be overcome before results can be visualized explicitly.  Fortunately, the phase can 
be retrieved by exploiting the causal properties of $\chi({\bf k},\omega)$, i.e. the fact that it satisfies 
the Kramers-Kr\"{o}nig (KK) relation, the second of which may be written\cite{jackson}

\begin{equation}
Re \left [\chi ({\bf k},\omega) \right ] = \frac{2}{\pi} P \int_{0}^{\infty} d\omega' 
\frac{Im \left [\chi ({\bf k},\omega) \right ]}{\omega' - \omega}.
\end{equation}

\noindent
Eq. (2) ensures that $\chi({\bf x},t)$ = 0 for $t < 0$ and 
is equivalent to choosing a retarded causality convention for the propagator.  
From (2) Re[$\chi({\bf k},\omega)$] may be determined, 
which in principle allows reconstruction of $\chi({\bf x},t)$\cite{martinCDW,circumvent}.

We used this procedure to image the density disturbance generated by a point perturbation in liquid water.  
The elementary quantum of density oscillation in water is the plasmon, which 
arises from collective vibration of the $2s$ and $2p$ shells of Oxygen whose density determines the 
normal frequency $\omega_p = 22$ eV\cite{kaoWater}. 
The plasmon is also the fundamental source of electronic 
screening in water and determines, for example, its optical refractive index\cite{nwater}.  
We used IXS to sample the plasmon momentum and energy dependence over 
the ranges 1 eV $< \omega <$ 100 eV and $0.476 < k < 4.95 \AA^{-1}$ (Figure 1).
Measurements were carried out on station C-1 at the Cornell High Energy Synchrotron Source 
(CHESS), where a nested Si(111)/(400) monochromator 
with sagittal focusing was employed\cite{ken}, and at CMC-CAT at the Advanced Photon Source (APS), where a
single Si(333) channel cut was used.  Both experiments used a diced, 
backscattering Ge(444) analyzer to achieve overall energy and momentum resolutions 
of $\Delta\omega = 0.30$ eV and $\Delta k = 0.146 \AA^{-1}$.
This configuration provided effective spacetime resolutions of
$\Delta t=41.3$ {\it as} and $\Delta x=1.27 \AA$, with fields of view of 13.8 {\it fs} and 54.3 $\AA$.  

In addition to inelastic scattering the spectra contained zero loss lines due to the static 
liquid structure factor.  This was subtracted and the spectra extrapolated 
to 0 at $\omega = 0$, where $Im[\chi({\bf k},\omega)]$ must vanish by causality, and at $k = 0$, 
where it must vanish by charge conservation.  
The $F$ sum rule was applied with $n_0 = 0.20 /\AA^3$ to place the spectra on an absolute scale.

While the proposed Kramers-Kr\"onig inversion procedure is simple in principle, unexpected 
subtleties arise when applying it to a discrete and finite data set.  
First, a discrete Im[$\chi(\omega)$] 
implies that $\chi(t)$ is periodic, a property incompatible with the causality constraint that 
$\chi(t)$ vanish for 
all $t < 0$.  Second, the KK relations are defined on an infinite $\omega$ interval, 
yet our scan range was a finite
100 eV.  We resolve the former issue by analytically continuing the spectra onto a continuous freqency 
interval, by linear interpolation, resulting in a time axis which is formally infinite.  
This permits $\chi(t)$ to vanish for all $t < 0$, but brings about an aliasing effect
that makes the dynamics repeat with a period 
$T = 2\pi/d\omega = 13.8$ {\it fs}, where $d\omega$ is the spectrometer resolution 
(here the quantity $2\pi/d\omega$ plays the role of 
the Nyquist critical frequency familiar from Fourier optics).  

The latter issue is resolved by extrapolating 
the spectra to $\omega \rightarrow \infty$ with a Lorentzian fit, which is appended 
to the measured spectra (Figure 2).  
This forces the time domain to be formally {\it continuous}, causing $\chi({\bf x},t)$ to be defined 
even on times much shorter than the effective  resolution of $2 \pi/100$ eV $= 41.3$ {\it as}.  It is
fair to examine the function on such short time scales, though
one must remember that its behavior is sensitive to the form of the extrapolation (Figure 2).  

Figure 3 displays time frames of $\chi({\bf x},t)$.  
Because the medium is isotropic the disturbance is spherically 
symmetric, and each image is a section of this sphere with $\chi({\bf x},t)$ 
plotted in units of $\AA^{-6}$ on the vertical 
axis.  At negative times $\chi({\bf x},t)$ vanishes and the system is "placid" (Fig. 3{\it a}).  
At $t = 0$ the system is 
"struck" with a positive perturbation, generating a negative recoil at the origin surrounded by a positive 
build-up at $|x| = 1.3 \AA$ as current flows away from the source point (3{\it b}).  
On an expanded scale a dip is 
also visible at $|x| = 2.5 \AA$, followed by another peak at $3.8 \AA$ (3{\it h}), 
forming a transient structure analogous
to the Friedel oscillations that occur around point impurities in metals\cite{friedel,commentFriedel}.  

Because of the compensating charge from the ion cores the pattern (3{\it h}) experiences a return force which 
causes it to reverse direction and change shape (3{\it c}).  An anharmonic oscillation occurs, with a time 
scale of order the plasma frequency $T = 2\pi/\omega_p = 180$ as (3{\it d-g}).  
Overall, the disturbance is seen to be rather local, occurring within 5 $\AA$ of the 
origin, a property related to the broad line shape in Figure 1.  A plasmon in water can oscillate a few 
times but never develops into a true propagating mode.  

As time evolves the entropy grows and the 
disturbance damps as the system evolves toward thermal equilibrium.  The disturbance decays below our
threshold of detectability, determined by the experimental noise level shown in Fig. 3{\it i}, 
after an elapsed time of 350 {\it as} $-$ less time than it takes light to travel 100 nm in vacuum.

$\chi({\bf x},t)$ provides a means to visualize electron dynamics, but more
significant is its relevance to a rather broad class of ultrafast processes in matter.
Specifically, because it describes the disturbance from a point source, 
by the principle of superposition $\chi({\bf x},t)$ 
can be used to determine the effects of extended sources.  
The charge $n_{ind}({\bf x},t)$ induced 
in a medium by a time-dependent source $n_{ext}({\bf x},t)$ is 
determined in reciprocal space by the relationship\cite{pines}

\begin{equation}
n_{ind}({\bf k},\omega) = \frac{4\pi e^2}{\hbar k^2} \chi({\bf k},\omega) n_{ext}({\bf k},\omega)
\end{equation}

\noindent
which in real space is a convolution over the dimensions of $n_{ext}({\bf x},t)$ identical 
to the Green's function integral used to handle sources in ordinary differential equations\cite{mathwalk}.  
We illustrate its use by modeling an oscillating dipole, and a gold ion travelling in water at 0.01 
times the speed of light, $c$.  As a model of the former we take 
$n_{ext}({\bf x},t) = [ \delta(x + 0.5 \AA) - \delta(x - 0.5 \AA) ] \cos(\omega_0 t)$, 
i.e. two point charges with opposite sign 
separated by 1 $\AA$, oscillating with a frequency $\omega_0 = 2$eV$/\hbar$.  
The resulting disturbance, $n_{ext}({\bf x},t)$, 
is shown in Figure 4a.  Such a disturbance would occur, for example, around a photoexcited chromophore in 
solution, and should have influence on short-range fluorescence resonance energy transfer (FRET) commonly 
used to measure distances in biological systems\cite{stryer}.  Similarly, an ion with charge $Z$ 
traveling at velocity {\bf v}, 
represented as $n_{ext}({\bf x},t) = Z \delta({\bf x} - {\bf v} t)$, 
for $Z=79$ and ${\bf v} = 0.01 c$ produces the pattern in Figure 4{\it b}.  This is an image 
of the density wake of a 9 MeV gold ion traveling in water 
and can be used, for example, to quantify the stopping power, dE/dx\cite{penn}.

It is important to address the issue of why, despite the spatial resolution of 
$dx = 1.27 \AA$, individual atoms are not visible in Figs. 3$-$4.  The reason is that a density propagator 
is actually a function of two spatial variables, $\chi({\bf x},{\bf x}';t)$, 
with a Fourier transform of the more 
general form $\chi({\bf k},{\bf k}';\omega)$.  There being only one momentum transfer in scattering, 
we measure only the diagonal (i.e. longitudinal) response, $\chi({\bf k},{\bf k};\omega)$\cite{sturm}. 
In other words the present analysis has assumed a translational invariance that does not strictly 
apply.  The images in Figs. 3$-$4 must therefore be thought of as spatial averages $-$ not single events 
but the mean of many events at different locations in the specimin.  In many cases an ensemble average 
provides adequate information; if not this limitation can be overcome by using
standing wave techniques\cite{schuelke2,golovchenko}.

The existence of nonlocality in electrodynamics, in both time and space, 
is well established\cite{jackson,agranovich}.  
In this article we have shown that they may be combined to explicitly 
image electron dynamics.  Because x-rays may be tuned over extremely broad ranges of energy,
time scales are accessible that are currently out of the reach of laser-based techniques.  
We therefore suggest
that IXS, because of its kinematic flexibility, may provide an alternative window
on the attosecond phenomena targeted in 
Refs. \cite{drescher1,drescher2,baltiuska,hentschel,keinberger,niikura1,niikura2,paul}.  A sensible
starting point for comparison would be the photofragmentation reaction of NaI or LiF.

Not all attosecond phenomena may be probed with IXS.  Like x-ray diffraction, it is only 
sensitive to phenomena 
which modulate the electron density, that is, which are capable of screening charge, 
such as collective modes like plasmons and phonons.  This makes IXS somewhat inappropriate for 
the study of low energy single particle excitations except in materials which cannot
sustain collective electronic vibration, such as large gap insulators\cite{wolfgang}.  
This is also true of resonance techniques which rely specifically on the coulomb interaction
to reach final states\cite{john,me,zahid}.
On the other hand, low energy electronic screening, 
which is determined by the plasmon through its real part, is quite accessible.  

We gratefully acknowledge experimental support from T. Gog, and helpful input from 
P. M. Platzman, N. W. Ashcroft, T. Brabec, G. Toombes, Wei Ku, I. K. Robinson, and S. K. Sinha.
This study was supported by the U.S. Dept. of Energy grant DEFG02-97ER62443.  
CHESS is supported by the National Science Foundation 
and the National Institute of General Medical Sciences under cooperative agreement DMR-0225180.  
The APS is supported by the U.S. Department of Energy under Contract W-31-109-ENG-38.

\begin{figure}
\caption{False colour plot of $-Im[\chi({\bf k},\omega)]$ for water in units of {\it as}/$\AA^3$, 
plotted against transferred momentum (horizontal axis, in $\AA^{-1}$) and energy (vertical axis, eV$^{-1}$).  
The broad feature is the valence plasmon, corresponding to collective oscillation of the $2s$ and $2p$ 
shells of oxygen.  
}
\end{figure}

\begin{figure}
\caption{Individual spectra at $k = 0.583 \AA^{-1}$ (above) and 4.95 $\AA^{-1}$  (below), 
showing the raw data (open circles), extrapolated tails (red lines) and the resulting real part (green) line.
}
\end{figure}

\begin{figure}
\caption{Time frames of $\chi({\bf x},t)$ derived from Figure 1, in units of $\AA^{-6}$.  
The vertical scale has been clipped at 1 $\AA^{-6}$ in frames (a-g), 0.1 $\AA^{-6}$ in (h), 
and 0.005 $\AA^{-6}$ in (i).  Distances are indicated with scale bars.  
(a) At $t < 0$, before the perturbation, the system is "placid". 
(b) Shortly after the initial event, showing a large (off scale) negative recoil at $|x| = 0$ 
surrounded by compensating positive build up. (c-g) The evolution of this disturbance at selected 
later times. (h) Same as (b) but on an expanded scale, showing distant Friedel-like oscillations. 
(i) After the disturbance has damped, scale expanded to show the experimental noise level.}
\end{figure}

\begin{figure}
\caption{Electron density disturbances for two extended sources, constructed via Eq. 3.  
(a) Induced charge around an oscillating dipole, such as a photoexcited chromophore in solution, and 
(b) the wake produced in water by a 9 MeV gold ion.  
The ion location and direction of motion are indicated by the red arrow.
}
\end{figure}

\end{document}